# GL-BLSTM: a novel structure of bidirectional long-short term memory for disulfide bonding state prediction

Junshu Jiang[1], Shangjie Zou[1], Yu Sun[1], Shengxiang Zhang[2,*1]

## Abstract

**Background:** Disulfide bonds are crucial to protein structural formation. Developing an effective method to predict disulfide bonding formation is important for protein structural modeling and functional study. Most current methods still have shortcomings, including low accuracy and strict requirements for the selection of discriminative features.

**Results:** In this study, we introduced a nested structure of Bidirectional Long-short Term Memory(BLSTM) neural network called Global-Local-BLSTM (GL-BLSTM) for disulfide bonding state prediction. Based on the patterns of disulfide bond formation, a BLSTM network called Local-BLSTM is used to extract context-based features around every Cys residue. Another BLSTM network called Global-BLSTM is introduced above Local-BLSTM layer to integrate context-based features of all Cys residues in the same protein chain, thereby involving inter-residue relationships in the training process. According to our experimental results,GL-BLSTM network performs much better than other methods, GL-BLSTM reached 90.26% accuracy at residue-level and 83.66% at protein-level. This model has reached the state of the art in this field.

**Conclusion:** GL-BLSTMs special structure and mechanisms are beneficial to disulfide bonding state prediction. By applying bidirectional LSTM, it can extract context based features by processing protein sequences, thereby obtain more discriminative information than traditional machine learning methods. Whats more, GL-BLSTMs special two-layer structure enables it to extract both local and global features, in which global features are playing important roles in improving prediction accuracy, especially at protein-level.

**Keywords:** Disulfide bonds; Prediction; Deep learning; Bidirectional long short term memory

## Background

Deep machine learning has been widely used in many research fields. Some machine learning methods such as artificial neural network (ANN) and support vector machine (SVM) are currently using in biological research, like processing clinical data for disease diagnosis and molecular classification [1, 2]. Natural organisms mainly use DNA and protein to store and convey information, while current methods like ANN and SVM are not suitable for handling biological sequences. The new architecture of neural network long short term memory(LSTM) has shown to have good performance for biological sequences. LSTM and
its variant called bidirectional LSTM (BLSTM) have achieved great successes in analyzing protein sequential data for secondary structure prediction and subcellular localization [3, 4], which showed great potential in biological data mining.

Disulfide bonds, also known as SS-bonds, are bridges between the thiol groups of two peptidyl-cysteine residues. The formation of disulfide bond is an oxidative process that generates a covalent bond between two cysteine residues in the same protein molecule. Disulfide bonds can be divided into two groups: inter-chain and intra-chain bonds. Inter-chain bonds produce stable protein multimers, while intra-chain bonds contribute to protein folding and stability. Thus, disulfide bridges are crucial to the formation of tertiary and quaternary structure of proteins [5], and even determine proteins' functions and physical properties, such as thermodynamic stability [6]. Therefore, correct disulfide bonding prediction can benefit both protein structural and functional researches.

According to former studies, the prediction of disulfide bonds includes two stages. Stage I is disulfide bonding state prediction, which concentrates on predicting whether a Cys residue is involved in a disulfide bond formation or not. Based on stage I, the second stage aims to figure out disulfide connectivity patterns and estimate locations of disulfide bonds. In the past three decades, many machine learning algorithms have been measured on disulfide bonding states prediction.Normally, protein sequences will be encoded by features including physicochemical properties, evolutionary information, hydrophobicity, etc., and then fed into machine learning models. Martelli et al. developed a hybrid system called hidden neural network which combined hidden Markov model and standard feed-forward neural network, and reached 87.4% accuracy on residue-level and 80.2% on protein-level by using RD dataset [7]. Yaseen et al. introduced context-based features in their enhanced feed-forward back-propagation neural network model and reached 88.8% accuracy at residue-level and 75.1% at protein-level by using Cull25 dataset [8]. However,

---

*Correspondence: sxzhang@scau.edu.cn
[1] College of Life Sciences, South China Agricultural University, Guangzhou 510642, China
[2] College of Mathematics and informatics, South China Agricultural University, Guangzhou 510642, China

traditional methods have their limitations. Algorithms such as ANN and SVM cannot fully utilize information stored in sequences. Because of that, these methods emphasize the selection of features for protein sequence encoding. Precise and massive information which describes protein properties are required for model to achieve high accuracy.

However, proteins which are newly identified and sequenced may not have sufficient information for prediction. Furthermore, the formation of one disulfide bond can influence statements of all Cys residues in the same protein molecule by affecting the overall structure of the protein[9]. However, previous methods could not considered this holistic effect efficiently. To overcome these defects, we proposed a novel structure of neural network system, called GL-BLSTM. It consists of two bidirectional LSTM (BLSTM) layers: Global-BLSTM layer and Local-BLSTM layer. Local-BLSTM layer processes "local" subsequence around each Cys residue, while Global-BLSTM layer can globally learn the relationships different between "local" subsequences by analyzing features extracted by Local-BLSTM. The double BLSTM layers can notably improve prediction accuracy and narrow the gap between residue-level and protein-level prediction. Furthermore, this model also has enormous potential in handling other difficult problems that require both local and global information such as disulfide connectivity prediction, which is still an unresolved problem [10].

## Methods

### Dataset

The training dataset is generated from PISCES server on 03/02/2018 [11]. The parameter setting for PISCES is shown below: pair-wise sequence identity is less than 25%; sequence length is between 40 and 10000; the cutoff of resolution is 3.0 and the R-factor value is 1.0. The raw training dataset has 13019 proteins.

The protein chains are filtered to fit our model. Protein chains with inter-chain disulfide bonds are abandoned. We exclude protein chains which contain less than 2 and more than 25 cysteine residues, to reduce computations for GL-BLSTM model. Through these filtering procedures, 7452 sequences are selected as final training dataset, which is named as Cull25-2018. Sequences of these proteins are processed by DSSP program [12] to assign the disulfide bonding positions. In total, Cull25-2018 contains 35921 Cys residues, with 18.97% are bonded.

### Feature extraction

*Cysteine-centered window and window size*

In this section, we demonstrate how to prepare the high-quality input for feature extraction. We applied cysteine-centered windows to slice subsequences from proteins' chains. A cysteine-centered window contains a Cys residue at its center, and (window size−1)/2 residues on upstream and downstream side of the central Cys residue (e.g. when window size is 7, the subsequence will contain one Cys residue at its center, and 3 other residues on each side). Each residue inside cysteine-centered window is then encoded with a matrix. The encoding matrix for each residue consists of 20 values from PSSM, 2 values of positional index and 2 values of physicochemical properties (Figure 1). Window size for GL-BLSTM is set as 7 according to our experiments and former studies [8, 13].

*Position-specific scoring matrix (PSSM)*

Position-specific scoring matrix (PSSM) can introduce evolutionary information in proteins encoding. It has been shown to significantly improve the overall prediction accuracy of disulfide bonding state prediction [14]. PSSM for protein samples in this project is generated by searching against Swiss-Prot database through three iterations of PSI-BLAST with a cut-off E-value of 0.001. For a protein sequence with L amino acid residues, PSSM can transform it into a L×20 matrix:

$$PSSM = [P_{i1}, P_{i2}, ..., P_{ij}, ..., P_{i20}] \quad (1)$$
$$(i = 1, 2, 3, ..., L; j = 1, 2, 3, ..., 20)$$

Eq.(1) shows the PSSM matrix for the i th residue in protein chain sequence(i = 1,2,3,...,L). The element $P_{ij}$ in this matrix indicates the score of i th residue in the protein sequence being mutated to amino acid type j. As there are twenty common amino acids, j ranges from 1 to 20. The values of $P_{ij}$ are directly generated by PSI-BLAST. After generating PSSM for residues, those PSSM values should be standardized by Eq.(2):

$$P_{ij}' = \frac{(P_{ij} - \sum_{k=1}^{20} P_{ik})}{(P_{max} - P_{min})} \quad (2)$$

where $P_{ij}'$ is standardized $P_{ij}$ value. $P_{max}$ is the maximum $P_{ij}$ value for the corresponding residue, while $P_{min}$ is the minimum one.

*Physicochemical properties*

Residues' physicochemical properties are included in the encoding matrix, as these factors improves machine learning models and disulfide bonding prediction [15]. Residues' polarity and hydrophobicity are selected because they are related to protein chains' re-gional structures. Data of these properties are obtained from previous researches [16, 17] and normalized by Eq. (3), in which X is the normalized value for polarity or hydrophobicity, while mean(x), max(x) and min(x) are representing the average, maximum and minimum values for the twenty residue types respectively.

$$X = \frac{x - mean(x)}{max(x) - min(x)} \quad (3)$$

*Residue's position*

Residues' positional information is also useful in this task [18]. In our model, each residue's position is represented by two values, which illustrate the distance from that residue to N-terminal and C-terminal of the protein chain respectively. Eq. (4) and Eq. (5) demonstrates how to calculate these two values. For the i th residue counted from the N-terminal of the protein, its position is demonstrated by DTN (Distance to N terminal) and DTC (Distance to C terminal). L is the length of protein chain that the residue belongs to. For i = 1,2,...,L,

$$DTN = (i - 0)/10000 \quad (4)$$
$$DTC = (L - i)/10000 \quad (5)$$

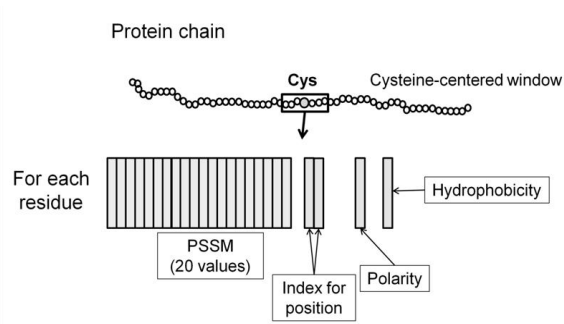

**Figure 1: Encoding of cysteine-centered window.** Protein subsequences centered by Cys are encoded with matrices. Each matrix contains data described above (20 columns of PSSM, 2 columns for positional index, 1 column for index of polarity and 1 for hydrophobicity).

*Protein subsequence encoding*

Feature information mentioned above enables us to encode each residue into a 1×24 matrix, with 20 columns of values from PSSM, two columns of positional index, one column for residue's polarity and one column for hydrophobicity (Figure 1). In this way, each cysteine centered window will be encoded with a [(window size)×24] matrix. If the window overlaps N-terminal or C-terminal, the window space outside the protein's chain will not produce PSSM values and physicochemical properties, but the positional indexes will still be assigned.

## "Force to Even" regulator

In this study, we apply the "Force to even" regulator in GL-BLSTM. The principle of this regulator is: The number of bonded cysteines must be even in a protein. When the number of Cys residues being predicted as "bonded" in one protein is odd, the system will reverse the prediction of the most "uncertain" residue and force the number of bonded Cys residues to be even. The most "uncertain" residue means the Cys residue whose probabilities of "Bonded" and "Free" are closer than others in that protein. Several studies [7, 19] have applied such strategy to improve the accuracy of prediction.

## ANN

We measured the performance of standard feed-forward artificial neural network (ANN) in this study. Our ANN model also used cysteine-centered window to slice subsequences for encoding. We have applied different window size for ANN to figure out its optimal performance. ANN is used as baseline in this study.

## Bidirectional LSTM (BLSTM)

Long short term memory (LSTM) [20] is a hot topic in deep learning. It is a variant of recurrent neural network (RNN). RNN builds connections between hidden units and utilize feedback loops to maintain information over time. Such features allow RNN to handle sequential inputs with arbitrary length and generate a fixed length vector. However, RNN suffers from gradient vanishing and exploding when processing long sequences [21]. RNN with LSTM blocks is developed to settle this problem, due to the fact that LSTM block can maintain information for a long period. Structure of LSTM memory block is shown in Figure 2.

In a LSTM block, three gates control the information flow. Through the coordination of input gate and forget gate, information is stored or vanished from the memory cell. Output gate controls the output of hidden state ($h_t$). Sigmoid function is used to implement these gates. By producing vectors with values ranged from 0 and 1, these sigmoid gates can filter information through element-wise multiplication with values squashed by tanh function. The values generated by these sigmoid gates depend on the hidden state of the last block ($h_{t-1}$) and the new input ($X_t$).

The following equations (6)-(11) state the mechanisms of LSTM units. In these formulations, σ is the symbol of sigmoid function, which produces values bounded by 0 and 1. ⊙ denotes element-wise multiplication, while it, ft and ot represents input gate, forget gate and output gate at $t^{th}$ time step respectively. The modulation for input squashing at time step t is denoted as $s_t$, and c t is the internal cell state at time step t. W, U and b are parameters to be learned. The tanh function in st gate can squash inputs to values range from −1 to 1.

After input squashing, the input gate, which produces values from 0 to 1, can filter information that has flowed through the st gate by element-wise multiplication: $i_t$ ⊙ $s_t$ . Those inputs multiplied by 0 would be rejected, while those multiplied by 1 would be accepted by the gate respectively. Similar to input gate, the forget gate regulates the forget process and remove some of the information out of the memory cell through the element-wise multiplication $f_t$ ⊙ $c_{t-1}$ , in which $f_t$ is assigned by sigmoid and works as a filter. These two operations together update the internal cell state from $c_{t-1}$ to $c_t$ . The similar mechanism can also be seen at output gate. The mechanisms mentioned above enable LSTM to memory information for a long period and avoid gradient vanishing [21].

$$i_t = \sigma(W^{(i)}x_t + U^{(i)}h_{t-1} + b^{(i)}), \quad (5)$$
$$f_t = \sigma(W^{(f)}x_t + U^{(f)}h_{t-1} + b^{(f)}), \quad (6)$$
$$o_t = \sigma(W^{(o)}x_t + U^{(o)}h_{t-1} + b^{(o)}), \quad (7)$$
$$s_t = \tanh(W^{(s)}x_t + U^{(s)}h_{t-1} + b^{(s)}), \quad (8)$$
$$ct = i_t \odot s_t + f_t \odot c_{t-1}, \quad (9)$$
$$h_t = o_t \odot \tanh(c_t), \quad (10)$$

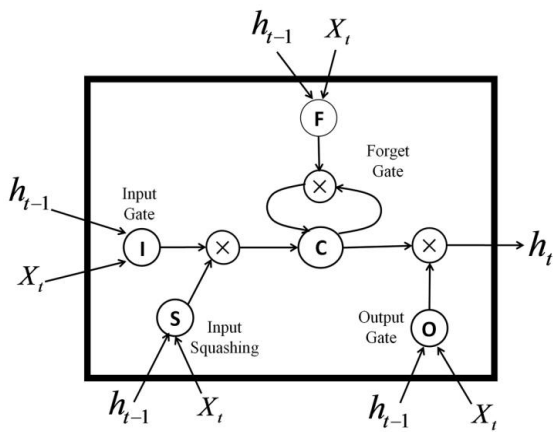

**Figure 2: Structure of LSTM memory block.** The three gates, input gate, output gate and forget gate, are denoted as I, O and F respectively. These gate are influenced by hidden state of the last block ($h_{t-1}$) and the new input ($X_t$). C is the memory cell and S is the node of Tanh function for input squashing.

Bidirectional LSTM [22] is optimized LSTM, which can read input sequences from both ends. This structure enables LSTM to learn sequential patterns from both directions. After processing, each time step t of BLSTM will generate two hidden states: $h_t^f$ (hidden state for forward layer at time step t) and $h_t^b$ (hidden state for backward layer at time step t). BLSTM models can be divided to two types in terms of the way they output their hidden states: many-to-many process and many-to-one process. For many-to-many process, the model would concatenate $h_t^f$ and $h_t^b$ and output a statement on every time step. By contrast, models for many-to-one process would only generate one statement, which is determined by the hidden states produced by the two directional layers on the last time step. For those situations which require an output on every time step, many-to-many structure is more suitable. On the contrary, when the situation requires the model to produce one generalized output, many-to-one structure would be prior. In this work, BLSTM neural networks are used as components to assemble GL-BLSTM, in which Local-BLSTM layer is many-to-one structure, and Global-BLSTM layer is many-to-many structure. We have measured the performance of BLSTM so as to make comparisons with GL-BLSTM. Our BLSTM model shares the same encoding strategy as ANN. The most suitable window size for BLSTM model is also worked out by experiment, and its best performance is used to compete with GL-BLSTM.

### Neural network architectures of GL-BLSTM
This section introduces the architectures of GL-BLSTM. BLSTM is efficient in handling sequences. However, formation of disulfide bond at one site is not only related to the Cys residues' local sequential environment, but also determined by the statements of other Cys residues and inter-residue relationships within one polypeptide chain (distance between one another, etc.) [23]. Single layer of BLSTM (so called Local-BLSTM layer) can only process sequences separately. By adding another layer of BLSTM (so called Global-BLSTM layer), our model can integrate features extracted from all Cys residues within one protein's chain. As shown in Figure 3, GL-BLSTM consists of four layers: input layer, Local-BLSTM, Global-BLSTM, time distributed output layer. The input layer transfers each residue among the cysteine-centered window into a 1 × 24 matrix, as described in feature extraction, and then feeds it into the Local-BLSTM layer. Size of cysteine-centered window for GL-BLSTM is set as 7 according to the experimental results and former studies [8, 13]. The Local-BLSTM layer aims to process the matrices for cysteine-centered subsequences and extract the sequence patterns. The state generated by Local-BLSTM layer would be fed to Global-BLSTM for further processing. The Global-BLSTM layer is stacked above Local-BLSTM layer. It can integrate the hidden states produced by Local-BLSTM Layer and utilize the patterns of these hidden states to optimize the training. Finally, each time step will output two numeric values at time distributed output layer. These two values represent the probabilities of the corresponding Cys residue being "bonded" or "free". Based on the values of two probabilities, our model will determine the state of Cys residue.

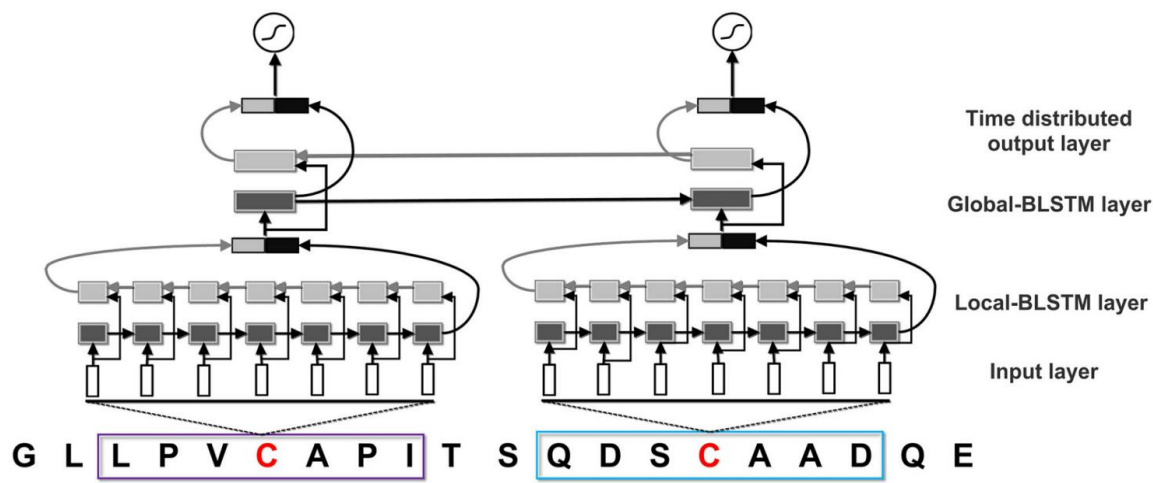

**Figure 3: Architecture of GL-BLSTM.** Input layer transfers each residue in cysteine-centered window to a 1 × 24 matrix. In this way, every cysteine-centered window can be transfered to a 7 × 24 matrix to represent the Cys residue at the center of the subsequence. Each node at input layer represents matrix of one residue. Matrices are then fed into Local-BLSTM layer. Local-BLSTMs correspond to different windows are separated. Features extracted from local windows are fed into Global-BLSTM. Finally, the network outputs the prediction of bonding statement for each Cys, which is represented by two values with probabilities of that Cys residue to be bonded [P(Bonded)] or free [P(Free)].

*Input layer*

Input layer transfers sequence region among cysteine-centered windows to 7 × 24 matrices. These representing matrices will then be fed into Local-BLSTM. Each node in Local-BLSTM corresponds to one residue. The size of cysteine-centered window is set as 7 based on the results of our experimental results and former studies [8, 13].

*Local-BLSTM layer*

In our GL-BLSTM model, Local-BLSTM layer can be seen as feature extractors for Cys residues' local environment. Pattern extracted by this layer will be passed to Global-BLSTM for final prediction. As shown in Figure 4, Local-BLSTM is a many-to-one BLSTM neural network. Local-BLSTM for each cysteine-centered window consists of two directional LSTM layers (forward and backward), each LSTM layer has 7 memory blocks, corresponds to 7 residues in each cysteine-centered window. In our study, Local-BLSTM only output the hidden layer state of the last time step on each direction. Then forward and backward hidden state are concatenated. After processing, Local-BLSTM will generate one hidden state to represent each cysteine-centered window.

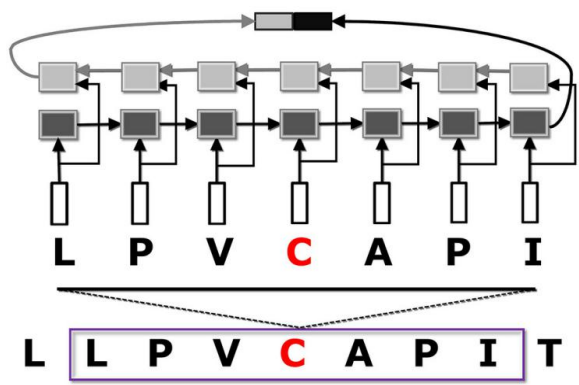

**Figure 4: Local-BLSTM layer.** Each Residue within cysteine-centered window is encoded with 1×24 matrixas input. One Local-BLSTM neural network generates one concatenated hidden state.

*Global-BLSTM layer*

In Global-BLSTM layer (Figure 5), features extracted from the same protein by Local-BLSTMs will be analyzed together in one BLSTM network. As shown in Figure 5, Global-BLSTM is a many-to-many BLSTM neural network. For each time step, Global-BLSTM will return a concatenated hidden state. In this way, the inter-residue relationships will be effectively learned and facilitate the prediction for every Cys residue.

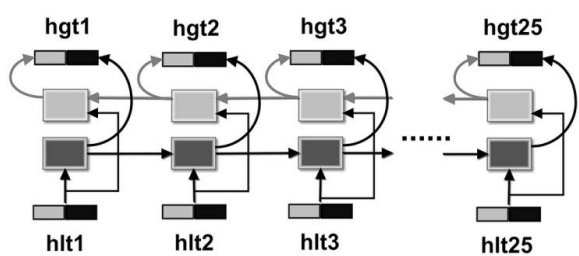

**Figure 5: A brief structure of Global-BLSTM layer.** Inputs of Global-BLSTM are the hidden states produced by Local-BLSTM layer, which is represented by hlt. After processing, each time step of Global-BLSTM will return a concatenated hidden state, which is denoted as hgt. t for the time step. The total number of GL-BLSTM time steps for each protein sequence is set as 25.

*Time distributed output layer*

Time distributed layer is a wrapper, which conduct output for every temporal node in Global-BLSTM layer. Each node in output layer will generate a binary vector to demonstrate the probabilities of bonding states. The binary vector consists of the probabilities of two bonding statements(bonded or free) for the corresponding residue.. The statement with higher probability will be selected as the prediction. This prediction is based on the hidden state of the corresponding Global-BLSTM time step.

*details for implementation:*

GL-BLSTM is implemented by Keras 2.1.3 with Tensorflow 1.4.0[24] as its backend. The number of hidden units of two BLSTM layers are set as 30. The activation function of the two BLSTM layers are ReLu [25]. The activation of the output layer is softmax and the loss function is cross entropy. The maximum sequence length of Local-BLSTM is set as 7 and the maximum sequence length of Global-BLSTM is set as 25. The model is optimized by Adam algorithm [26]. Batch normalization is adopted between Global-BLSTM layer

# Results

## Performance Measurement

The performance of our models were measured by 10 fold cross-validation. To conduct this validation, the dataset was randomly divided into ten parts with equal size. One part was used as test set and the others for training. To avoid systemic error caused by distribution of subsets, such process should be repeat for 10 times. In this way, 10 test sets and 10 training sets were prepared. We used sensitivity (*Sn*), specificity (*Sp*) and Matthews correlation coefficient (*MCC*) to measure the performance of cysteine state prediction. Definitions of these metrics are shown as follows:

$$Sn = \frac{TN}{TP+FN}$$

$$Sp = \frac{TN}{TN+FP}$$

$$MCC = \frac{TP \times TN - FP \times FN}{\sqrt{(TP+FN)(TP+FP)(TN+FP)(TN+FN)}}$$

In these equations, TP, TN, FP and FN refers to true positive, true negative, false positive and false negative respectively. $Q_c$ and $Q_p$, which represent accuracy at residue-level and protein-level respectively were widely adopted by former studies to evaluate models performance[10]. Eq. (12) and Eq. (13) demonstrates how these two metrics are calculated:

$$Qc = \frac{N_c}{N_b} \quad (11)$$

$$Qp = \frac{N_{prot}}{N_{total\_prot}} \quad (12)$$

In Eq. (12), $N_c$ indicates number of disulfide bonds that are correctly predicted, while $N_b$ is the total number of disulfide bonds in the test set. In Eq. (13), $N_{prot}$ is the number of proteins whose Cys residues bonding states are all correctly predicted, and $N_{total}$ is the total number of proteins in the test set.

## Window size for GL-BLSTM

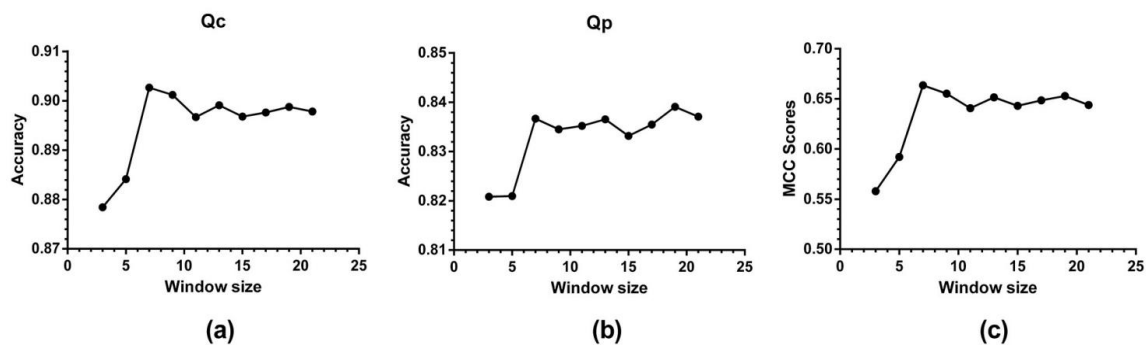

**Figure 6: Performance metrics of GL-BLSTM based on different window size.** (a) Accuracy at residue-level($Q_c$). (b) Accuracy at protein-level($Q_p$). (c) MCC scores for GL-BLSTM.

Based on the residue's information and BLSTM neural network, we constructed GL-BLSTM system for disulfide bonding state prediction.We first measured performance of GL-BLSTM on the basis of Cull25-2018. Figure 6a and Figure 6b show the accuracy of GL-BLSTM at residue-level and protein-level with different window size respectively. The sequential information fed to this two model was generated by the encoding strategy (each residue is represented by 1×24 matrix). GL-BLSTM model achieved its highest MCC score and the accuracy at residue-level when window size is 7. At this window size, GL-BLSTM achieved 90.26% Qc and 83.66% Qp , outperformed the former models that used the same strategy as ours to generate datasets.

Similar trend can also be seen in other metrics: sensitivity (Sn) and specificity (Sp), both of which reached a peak when window size is 7 (Figure 6c). Based on these figures, we decided to use 7 as the optimal window size. This pattern can be explained by former studies [8, 13], which have revealed that 7 residues can involve important residue correlations and avoid accumulation of sampling noise.

We have also noticed that when window size was larger than 11, the accuracy kept on fluctuating at a stable level overtime. Although more information had been input, and time consumption for calculation has grown excessively, there was no significant decrease or increase in predicting performance. This phenomenon is an outcome of the normalization and dropout methods [27, 28].

Performance comparison of proposed GL-BLSTM with other models

| Method | Qc (%) | Qp (%) | Dataset | Year | Reference |
| --- | --- | --- | --- | --- | --- |
| HNN (a) | 87.4 | 80.2 | RD | 2002 | Martelli et al. 2002 [7] |
| Dinosolve-ANN (b) | 88.8 | 75.1 | Cull25 | 2013 | Yaseen et al. 2013 [8] |
| ANN (c) | 85.31 | 70.56 | Cull25_2018 | 2018 | This work |
| BLSTM | 87.44 | 76.95 | Cull25_2018 | 2018 | This work |
| GL-BLSTM (Window size 7) | 90.26 | 83.66 | Cull25_2018 | 2018 | This work |

**Table 1. Accuracy of disulfide bonding state prediction between this work and previously published methods:** (a) In the study of Martelli [7], two datasets were used: 'WD' and 'RD'. 'WD' is 'whole dataset' and RD is 'reduced dataset' which reduced chain containing only one Cys residue. We removed chains with only one Cys residue as many other studies, and only compare our performance with the results obtained on RD dataset. (b) Yaseen et al. developed "Dinosolve" based on ANN and obtained good results (c) ANN is used to provide a baseline for measurements of BLSTM and GL-BLSTM models performance. It's notable that they generated their PSSM by searching against non-redundant database on NCBI, which is much bigger than Swiss-prot. This may be the reason why their ANN reached relatively high accuracy.

Table 1 compares GL-BLSTMs performance with ANN and BLSTM constructed in this study and two previously pioneered models: Dinosolve-ANN and HNN. According to the data, GL-BLSTM's Qp is 83.66%, much higher than the other models, which are mostly under 80%. Meanwhile, its accuracy at residue-level (90.26%) is also the highest one among all predictors.

## Discussion

Among the three neural network models constructed in this study, ANN can only utilize evolutionary and residue features, which are input in terms of matrices, while normal BLSTM can extract context based features. As a consequence, BLSTM overwhelmed ANN at both levels. For GL-BLSTM's prediction is based on both the local sequential environments and overall patterns in proteins, it can significantly improve the accuracy and narrow the gap between Qc and Qp . The accuracy of GL-BLSTM was 4.95% at residue-level and 13.1% at protein-level higher than ANN, while the gap between Qc and Qp was only 6.6%, much lower than both ANN and normal BLSTM. Furthermore, BLSTM reached its peak (Qc =87.44%, Qp =76.95%) when input window size is 73, while GL-BLSTM can perform much better, especially on protein-level, with less input (when window size is 7). Because the only difference between the two models is the addition of Global-BLSTM layer, it is clear that this layer has the capability of extracting and utilizing some potential discriminative features, like the correlations between different Cys residues bonding statements.

Looking at the experimental data for more details. In comparison with normal BLSTM, the GL-BLSTM can not only improve the prediction accuracy, but also narrow the gap of predicting accuracy between residue-level and protein-level. These phenomena illustrate that Global-BLSTM layers can improve protein-level accuracy by establishing a feedback pathway between Cys residues. Such a feedback pathway enables every cysteines predicted bonding statement to contribute to the anticipations of other residues. Therefore, it provides an efficient approach to train and optimize models weights, results in more accurate predictions.

## Conclusion

In this study, we propose a new end-to-end method called GL-BLSTM in disulfide bonding state prediction based on proteins' sequences. According to our experimental results, GL-BLSTM is one of the best systems for this task so far. The GL-BLSTM architecture is an assembled BLSTM neural network. Compares to normal BLSTM and ANN, this new architecture takes the relationships of different Cys residues in the same protein into account. According to our experiments, this model has successfully achieved 90.26% accuracy at residue-level and 83.66% at protein-level on the basis of Cull25-2018, much higher than the other pioneered models. Such a great achievement demonstrates that the inter-correlations between Cys residues should not be neglected.

The good performance of GL-BLSTM also illustrates that there are important correlations between different Cys residues, which can affect these residues disulfide bonding statements. This new discovery has revealed a new direction for further improvements of disulfide bonding prediction.

Apart from disulfide bonding state prediction, the architecture of GL-BLSTM may also benefit other applications. In the future, we may discover GL-BLSTMs effectiveness in handling other tasks, such as disulfide connectivity prediction.

# List of Abbreviations

ANN: Artificial Neural Network; BLSTM: Bidirectional Long Short Term Memory Neural Network; GL-BLSTM: Global-Local-Bidirectional Long Short Term Memory Neural Network; PSSM: Position-specific Scoring Matrix; RNN: Recurrent Neural Network; SVM: Support Vector Machine

# Declaration

### Ethics approval and consent to participate
Not applicable

### Consent to publish
Not applicable

### Availability of data and materials
The dataset and the source codes supporting the conclusions of this article are available in thegithub repository (https://github.com/scaujjs/GL-BLSTM)

### Competing interests
The authors declare that they have no competing interests.

### Funding
This work is supported by Science and Technology Program of Guangzhou (No.201707010031).

### Authors' contributions
Junshu Jiang and Shengxiang Zhang designed the study and performed the testings, Shangjie Zou and Yu Sun wrote the manuscript andassisted in analyzingthe results and optimizing the neural network model. All authors read and approved the final manuscript.

### Acknowledgements
We would like to thank all the members of mathematical laboratory, College of Mathematics and Informatics, South China Agricultural Universityfor their additional support. We also like to thank Mr. Bin Wen for his support in maintaining experimental facilities.